\documentclass[10pt,journal,compsoc]{IEEEtran}
\ifCLASSOPTIONcompsoc
\usepackage[nocompress]{cite}
\else
\usepackage{cite}
\fi
\usepackage{amssymb}
\usepackage{graphicx}
\usepackage{subfig}
\usepackage{changepage}
\usepackage{bm}
\usepackage[table,xcdraw]{xcolor}
\usepackage{color, soul}
\usepackage{makecell}
\usepackage{url}
\urldef{\mailsa}\path|{alfred.hofmann, ursula.barth, ingrid.haas, frank.holzwarth,|
	\urldef{\mailsb}\path|anna.kramer, leonie.kunz, christine.reiss, nicole.sator,|
	\urldef{\mailsc}\path|erika.siebert-cole, peter.strasser, lncs}@springer.com|

\usepackage{amsmath}
\usepackage{comment}
\usepackage{epsfig}
\usepackage{multirow}
\usepackage{booktabs}
\usepackage[noend]{algpseudocode}
\usepackage{algorithmicx,algorithm}
\usepackage{diagbox}
\usepackage{bbding}
\usepackage{textcomp,booktabs}
\usepackage{colortbl}
\usepackage{enumitem}

\usepackage{multirow}
\usepackage[normalem]{ulem}
\useunder{\uline}{\ul}{}

\begin{document}

%	\title{An Intent Disentangled and Self-supervision Enhanced Framework for Novel Recommendation}
	\title{Intent Disentanglement and Feature Self-supervision for Novel Recommendation}
	
	%\begin{comment}
	\author{Tieyun~Qian,~\IEEEmembership{Member,~IEEE,} Yile~Liang, Qing~Li,~\IEEEmembership{Senior Member,~IEEE}, 
		
	Xuan~Ma, Ke~Sun, Zhiyong~Peng,~\IEEEmembership{Member,~IEEE,}% <-this % stops a space
	\IEEEcompsocitemizethanks{\IEEEcompsocthanksitem T. Qian, Y. Liang, X. Ma, K. Sun and Z. Peng are with Wuhan University, China. \protect E-mail: qty@whu.edu.cn, liangyile@whu.edu.cn, 2017301500248@whu.edu.cn, sunke1995@whu.edu.cn, peng@whu.edu.cn
	% note need leading \protect in front of \\ to get a newline within \thanks as
	% \\ is fragile and will error, could use \hfil\break instead.
	\IEEEcompsocthanksitem Q. Li is with the Hong Kong Polytechnic University, Hong Kong. E-mail: qing-prof.li@polyu.edu.hk
	}% <-this % stops an unwanted space

	%\thanks{Manuscript received May 10, 2020; revised,.}
	}
	%\end{comment}
	
	% The paper headers
	\markboth{Journal of \LaTeX\ Class Files,~Vol.~14, No.~8, August~2015}%
	{Shell \MakeLowercase{\textit{et al.}}: Bare Demo of IEEEtran.cls for Computer Society Journals}
	
	\IEEEtitleabstractindextext{%
		\begin{abstract}

One key property in recommender systems is the long-tail distribution in user-item interactions where most items only have few user feedback. Improving the recommendation of tail items can promote novelty and bring positive effects to both users and providers, and thus is a desirable property of recommender systems.
%Which, however, is a challenging task due to the conflict between accuracy and novelty. %since enhancing novelty often reduces accuracy.
Current novel recommendation studies over-emphasize the importance of tail items without differentiating the degree of users' intent on popularity and often incur a sharp decline of accuracy. Moreover, none of existing methods has ever taken the extreme case of tail items, i.e., cold-start items without any interaction, into consideration.

In this work, we first disclose the mechanism that drives a user's interaction towards popular or niche items by disentangling her intent into conformity influence (popularity) and personal interests (preference). We then present a unified end-to-end framework to simultaneously optimize accuracy and novelty targets based on the disentangled intent of popularity and that of preference. We further develop a new paradigm for novel recommendation of cold-start items which exploits the self-supervised learning technique to model the correlation between collaborative features and content features.
We conduct extensive experimental results on three real-world datasets. The results demonstrate that our proposed model yields significant improvements over the state-of-the-art baselines in terms of accuracy, novelty, coverage, and trade-off.			
		\end{abstract}
		
		% Note that keywords are not normally used for peerreview papers.
		\begin{IEEEkeywords}
			recommender systems, novel recommendation, disentangled representation, self-supervised learning.
	\end{IEEEkeywords}}

	% make the title area
	\maketitle

	\IEEEdisplaynontitleabstractindextext
	
	\IEEEpeerreviewmaketitle
	
	\begin{comment}
	\IEEEraisesectionheading{\section{Introduction}\label{sec:introduction}}
	
	\IEEEPARstart{T}{his} demo file is intended to serve as a ``starter file''
	for IEEE Computer Society journal papers produced under \LaTeX\ using
	IEEEtran.cls version 1.8b and later.
	% You must have at least 2 lines in the paragraph with the drop letter
	% (should never be an issue)
	I wish you the best of success.
	
	\hfill mds
	
	\hfill August 26, 2015
	
	\subsection{Subsection Heading Here}
	Subsection text here.
	
	% needed in second column of first page if using \IEEEpubid
	%\IEEEpubidadjcol
	
	\subsubsection{Subsubsection Heading Here}
	Subsubsection text here.
	
	\end{comment}

	\section{Introduction}\label{sec:intro}
With the rapid growth of the web, the users are overwhelmed with the choice of ``finding the right thing'' from a vast number of products.
The recommender systems, which use historical data to infer the user's preference on particular items like movies, commodities, and places, are a crucial component of many e-commerce platforms.
One key property in recommender systems is the long-tail distribution in user-item interactions, where a tiny amount of popular items receive most of the user attention and a large proportion of tail items only have few user feedback. Improving the recommendation of tail items can enrich users' experience by providing them with more chances to find interesting yet unpopular items. It can also bring positive effects to the  providers since the niche products may increase the companies' marginal profits.  As a result, novel recommendation becomes a desirable property of modern recommender systems. Which, however, is a challenging task due to the conflict between accuracy and novelty.% i.e., enhancing novelty often reduces accuracy.

Most of existing studies on novel recommendation~\cite{AdomaviciusK_tkde12,YinCLYC_vldb12,Park_tkde13,Oestreicher-SingerS_misq12,Generic_icde18}, also known as the long-tail recommendation in the literature~\footnote{We will use the terms ``novel recommendation'' and ``long-tail recommendation'' to describe the existing work interchangeably.}, adopt a two stage re-ranking strategy. At the first stage, the recommender systems aim to achieving a high accuracy using a base model.  At the second stage, the results from the previous stage are re-ranked towards the novelty target by introducing more tail items into the candidate top-$N$ item list. Such approaches have an inherent limitation, i.e., the popularity bias that recommenders emphasize popular items  much more than tail ones remains exist at the first stage.
%Due to the existence of popularity bias~\cite{popbias_recsys17}, these systems tend to recommend popular items.

Several recent studies~\cite{MatchNov_icdm19,tailnet_recsys20,MIRec_www21} propose the end-to-end framework where the recommendation is optimized for both the accuracy and  novelty at the same time. While making progress, the PPNW and TailNet methods in ~\cite{MatchNov_icdm19,tailnet_recsys20} tend to recommend tail items and result in a relatively low accuracy. More importantly, none of the re-ranking and the end-to-end based novel recommendation methods differentiate the degree of users' intent to popularity. Indeed, though the consumers often refer to others for product choice and will comply with the group norm to some extent, an individual's conformity is affected by many factors like intelligence, personality, and status~\cite{Lascu_1999,Park_2010}. That is to say, facing the same popular item, different consumers may react in the opposite direction. Some users tend to conform the others' action and purchase the popular item, while others may or may not undertake the same action as they are more likely to  meet their own personal interests.

	\begin{figure*}[htb]
		\centerline{\includegraphics[width=0.9\textwidth]{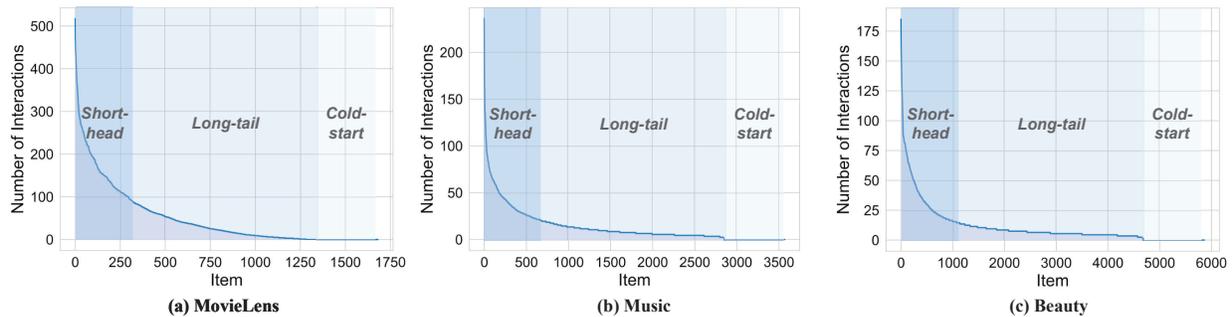}}
		\caption{Long-tail distribution of the items' interaction frequencies on three datasets.}
		\label{fig-case-longtail}
	\end{figure*}

In view of the different levels of conformity among users, we propose to disentangle a user's  intent into conformity influence and personal interests, which drives her interaction towards popular and novel niche items, respectively.  By doing this, the users' great diversity of intent can be uncovered to enhance the expressiveness of users' representations. The same operation is performed in the item side, i.e., disentangling items' representation into  popularity and characteristic factors.
We then present a unified end-to-end framework to simultaneously optimize accuracy and novelty targets based on the learned user and item representations. In this way, the intrinsic popularity and novelty factors are  introduced into the novel recommendation process and the systems can naturally achieve a balance between the recommendation of tail items and that of popular items.

We further make in-depth analyses on the relationship between the normal long-tail items and the cold-start ones. As shown in Figure~\ref{fig-case-longtail}, the cold-start items are actually the extreme case of long-tail items which do not have any interactions.
Kapoor et al. ~\cite{ilike_recsys15} point out in their pioneering work that an item can be novel in three ways: (1) it is new to the system and thus for every user (cold-start), (2) it is known to the system but new to the single user, (3) it is known to the user long before but forgotten at the moment.
Unfortunately, existing studies~\cite{AdomaviciusK_tkde12,YinCLYC_vldb12,Park_tkde13,Oestreicher-SingerS_misq12,Generic_icde18,MatchNov_icdm19,tailnet_recsys20,MIRec_www21} focus on novel recommendation tasks under the above definition (2) (3), and none of them has ever taken cold-start items into consideration. Meanwhile, previous research on cold-start recommendation ~\cite{GraphSAGE_NeuIPS17,IGMC_iclr20,2020-TKDE-FMFC,Aharon-Active-Recsys15,StarGCN_ijcai19,HERS_aaai19,AGNN_tkde20} does not include long-tail items, either.

Based on the above analyses, we develop a new paradigm for novel recommendation of cold-start items. To address the problem of missing collaborative features for cold-start items, we exploit two types of  self-supervised learning technique, including the variational autoencoder and mutual information maximization, to model the correlation between collaborative features and content features. To the best of our knowledge, this is the first attempt to exploring both the known long-tail  and the unknown cold-start items in novel recommendation.

Extensive experimental results on three datasets demonstrate that our proposed model yields significant improvements over the state-of-the-art baselines in terms of the overall trade-off among accuracy, coverage, and novelty metrics on novel recommendation task for both the standard long-tail and the cold-start items.

\begin{comment}
	Challenges:
	\begin{itemize}
		\item Due to the popularity bias~\cite{popbias_recsys17} and "rich gets richer" feedback loop~\cite{MaZYYCTHC_www20}, long-tail items are more difficult to be mined compared with short-head items. The users would interact popular items by reason of conformity effect, while they would prefer niche items derived from their intrinsic interest. How to disentangle user intents towards different items and estimate user preference further?
		
		\item The cold-start items have not any interactions, but they are equipped with inherent content information, e.g., attributes, textual descriptions and images. How to utilize the auxiliary content information to enhance the representation of cold-start items?
		
	\end{itemize}
\end{comment}

	\section{Related Work}
	In this section, we review the related work from three perspectives, namely novel recommendation, disentangled representation learning, and self-supervised learning.

	%\vspace{-1mm}
	\subsection{Novel recommendation}
%We partition research on novel recommendation into standard long-tail recommendation and cold-start recommendation.
	\textbf{Standard Long-tail recommendation.}
%Early recommender systems typically employ collaborative filtering approaches ~\cite{LFM_09,NCF_www17,CML_www17,cdae_wsdm16} for mining users' preference from their behaviors. Despite their success, these methods typically concern the most popular items and neglect the niche items highly relevant to the user. According to previous research~\cite{NovDiv_toit11,Generic_icde18,MatchNov_icdm19}, an item long-existing in the system but not popular to be seen by most users means novel. %As the long-tail distribution of user feedback is common in the recommender systems, its direct effect is the oblivion of a large number of tail items by the recommender.
	Current research on long-tail recommendation can be categorized into two groups. The first group design different re-ranking methods. They post-process the ranking list of a standard model to account for additional objectives like coverage rather than devising a new model. Some studies~\cite{AdomaviciusK_tkde12,Oestreicher-SingerS_misq12,Generic_icde18} improve novelty by countering the effects of item popularity, and others~\cite{Park_tkde13,YinCLYC_vldb12} propose clustering approaches and leverage tail items directly in a recommendation list.
	The second group of methods adopt end-to-end models. For example, Lo et al.~\cite{MatchNov_icdm19} propose a personalized pairwise novelty weighting for BPR loss function as an end-to-end method.  Liu and Zheng~\cite{tailnet_recsys20} present a network architecture for long-tail session-based recommendation by introducing an adjustable preference mechanism. Zhang et al.~\cite{MIRec_www21} transfer knowledge from head items to tail items for leveraging the rich user feedback in head items and the semantic connections between head and tail items.

Overall, both groups of long-tail recommendation models ignore the users’ different level of conformity. In contrast, we disentangle users' intent into popularity and preference embeddings, which capture the inherent factors that determine users' choice on popular or niche items.
	
	\textbf{Cold-start recommendation.} %Cold start is a fundamental and challenging problem in recommender systems. A promising approach to this problem is to leverage side information such as contextual information ~\cite{NFM_sigir17}, the user and item relations~\cite{RelationalCF_sigir19}.
Early cold-start recommendation methods ~\cite{TrustSVD_aaai15,NFM_sigir17,RelationalCF_sigir19} mainly exploit side information as regularization in MF objective function~\cite{TrustSVD_aaai15} or adopt similarity based or feature mapping technique for integrating side information.
More recent studies design various types of neural models to incorporate side information, including  graph-based inductive learning methods~\cite{GraphSAGE_NeuIPS17,IGMC_iclr20}, the model based meta-learning methods~\cite{2017-NeuIPS-LWA} and the gradient based meta-learning method  ~\cite{2019-SIGIR-MetaEmb}, the active learning scheme~\cite{2020-TKDE-FMFC,Aharon-Active-Recsys15} which require extra costs or budgets, and graph neural models~\cite{HERS_aaai19,AGNN_tkde20}.

While these methods achieve promising performance for cold-start recommendation, they are not designed for recommending long-tail items. For example, DropoutNet~\cite{DropoutNet_NeuIPS2017} and STAR-GCN~\cite{StarGCN_ijcai19} simulate  the cold-start scenario by masking and reconstructing a part of input features in training without changing the item representations. Such methods cannot be adapted to the novel recommendation. One of our contribution is to view the cold-start items as the extreme cases of tail items, and present a new learning paradigm for cold-start novel recommendation.
	
\begin{comment}
Several graph-based inductive learning methods~\cite{GraphSAGE_NeuIPS17,IGMC_iclr20} make the learned embeddings generalized to unseen users/items, but they need some links for cold-start nodes at the test phase.
	The model based meta-learning methods LWA and NLBA~\cite{2017-NeuIPS-LWA} and the gradient based meta-learning method  MetaEmb~\cite{2019-SIGIR-MetaEmb} can generalize the well-trained meta-learner to cold-start scenario.
	The active learning scheme~\cite{2020-TKDE-FMFC,Aharon-Active-Recsys15}, where a number of users are selected for rating or commenting on a new item, can also tackle this problem but with the extra costs or budgets.
	DropoutNet~\cite{DropoutNet_NeuIPS2017} and STAR-GCN~\cite{StarGCN_ijcai19} simulate and adapt the cold-start scenario when training by masking or reconstructing a part of input features.
	HERS~\cite{HERS_aaai19} and AGNN~\cite{AGNN_tkde20} aggregate user and item relations leveraging the social links or attribute graph in the form of graph neural network.
\end{comment}
	%\vspace{-1mm}

	\subsection{Disentangled Representation Learning}
	Disentangled representation learning is originated from computer vision field ~\cite{betaVAE_iclr17,HamaguchiSN_cvpr19}. It aims to  learn representations that separate explanatory factors of variations behind the data to improve the robustness and interpretability.
%For example, disentanglement learning on a visual dataset might learn the shape, the color, and the position features of the object, where each feature is not easily influenced when other feature changes. Lately, disentangled representation learning has been investigated on graph-structured data~\cite{DisenGCN_icml19}.

	With the superior performance, disentangled representation learning sheds new light on recommender systems. MacridVAE~\cite{MacroRec_nips20} is the first work that introduces the disentangled representation learning into user behavior data at both a macro and a micro level. Inspired by it, DICER~\cite{DisenCont_recsys20} combines the content information into the procedure of disentanglement. Besides, a few studies~\cite{DisenNews_acl20,DisenGCF_sigir20} apply disentangled graph convolutional networks to reflect the fine-grained latent intents. Ma et al.~\cite{DisenRec_kdd20} propose to disentangle the intents behind any given sequence of behaviors for sequential recommendation.

	All the aforementioned methods need to define $K$ latent intents/channels before performing disentangling and comprehend these latent representation through post-hoc explanation. Perhaps the closest work to ours is DICE~\cite{Zheng_Disentangling_www21} which contains user interest and conformity disentanglement. However, DICE is not for long-tail recommendation and does not need to consider the balance between accuracy and novelty. In addition, it adopts the casual model and negative sampling strategy to learn corresponding  embeddings. Different from it, we present an  intent prototype approach to aggregate users' and items' features, which tallies well with consumer conformity theory~\cite{Venkatesan_1966,Lascu_1999,Park_2010}.

	%By comparison, our work define the specific intents explicitly  and aim to identify the cause which drive a user interacts with different items, so as to dig out more relevant niche items to improve the novelty.
	
	\subsection{Self-supervised Learning}
	Self-supervised learning  models can leverage input data itself as supervision and benefits almost all types of downstream tasks. The objectives in self-supervised learning can be categorized into  generative, contrastive, and adversarial types~\cite{SSL_arxiv20} such as auto-encoding~\cite{dae_icml08,VAE_ICLR14}, VAE variants ~\cite{VGAE_nips16,VQVAE_nips19}, discriminative  models~\cite{PIRL_cvpr20}, and Adversarial self-supervised learning (GANs)~\cite{GANs_iclr16}.

	%Auto-encoding like denoising AE~\cite{dae_icml08} and variational AE~\cite{VAE_ICLR14} is the mainstream generative methods in self-supervised learning, which aims to reconstructing inputs from original inputs. More recently, many VAE variants have been employed in advancing research fields, such as image generation~\cite{VQVAE_nips19} and graph representation learning~\cite{VGAE_nips16}.Contrastive self-supervised learning belongs to discriminative model, which tries to discriminate the relationship of a paired example, e.g., predicting the relative position between parts of data samples~\cite{PIRL_cvpr20} or the association between a sample and its context~\cite{DIM_iclr19}. 	Adversarial self-supervised learning (GANs)~\cite{GANs_iclr16} utilizes contrastive objectives that distinguish the generated samples from real ones.
	
	As the research of self-supervised learning is still in its infancy, there are only several studies incorporating it with recommender systems~\cite{SGL_sigir21,MHCN_www21,DisenRec_kdd20,S3Rec_cikm20,BiGI_wsdm21,DHCN_aaai21}. Some of these efforts mine the self-supervision signals from sequential data~\cite{DisenRec_kdd20,S3Rec_cikm20,DHCN_aaai21} and others capture the structure properties from user-item bipartite graph~\cite{BiGI_wsdm21,SGL_sigir21} or social graph~\cite{MHCN_www21}. Different from the above approaches, our work is the first to consider the correlations between the collaborative features originated from interaction behaviors and user's/item's inherent content features, and we adopt the generative and contrastive self-supervised learning jointly under these two types of features to tackle the cold-start problem in recommendation.

	\section{Proposed Model}
	In this section, we introduce the proposed model by first formulating the problem and then presenting the details.
	
	\subsection{Problem Formulation}
	
	Let $U = \{u_1, u_2, ..., u_M\}$ be a set of users and $V = \{v_1, v_2, ..., v_N\}$ be a set of items, where $M$ and $N$ denote the corresponding cardinalities. Let $\bm{R} \in \mathbb{R}^{M \times N}$ be the user-item interaction matrix, indicating whether the user purchased or clicked on the item. We focus on the recommendation task with implicit feedbacks~\cite{HuKV08_icdm08}, where the interaction matrix is defined as:
	$$ R_{uv}=\left\{
	\begin{array}{l}
		1 \text{, if an interaction (user $u$, item $v$) is observed,} \\
		0 \text{, otherwise.}
	\end{array}
	\right.
	$$
	The observed entries reflect users' interest in the item, while the unobserved entries are mixed with unknown data and negative views of the item.
Given the above interaction information, our goal is to estimate users' interest for unobserved entries and rank the candidate items according to the predicted scores such that we can recommend the top-$N$ items for the traditional recommendation task.
	Beyond that, we are particularly interested in novel recommendation for both long-tail and cold-start items.
%According to the Pareto principle ~\cite{Pareto_86},

Recent studies show that the item space can be divided into two parts, namely popular short-head items and niche long-tail items. The long-existing but unpopular tail items have high novelty, and they can provide more surprises for users and more profits for providers. Besides, the cold-start items without any interaction can be regarded as extreme cases of tail items. The cold-start scenario is commonly found in the real world, e.g., the newly released movies have no audience. These new items can stimulate users' interest and avoid over-specialization in recommendation. Thus we extend the definition of standard long-tail recommendation to cover cold-start items.

The main target for novel recommendation is to optimize the trade-off between accuracy and novelty~\cite{Generic_icde18,MatchNov_icdm19,tailnet_recsys20}. On one hand, the recommendation accuracy is evaluated by a matching score between the recommendation list and the ground-truth list. On the other hand, the recommendation quality is measured by the proportion of novel items and the coverage of item space of the list.

	\subsection{An Overview}
	\begin{figure*}[htb]
		\centerline{\includegraphics[width=0.9\textwidth]{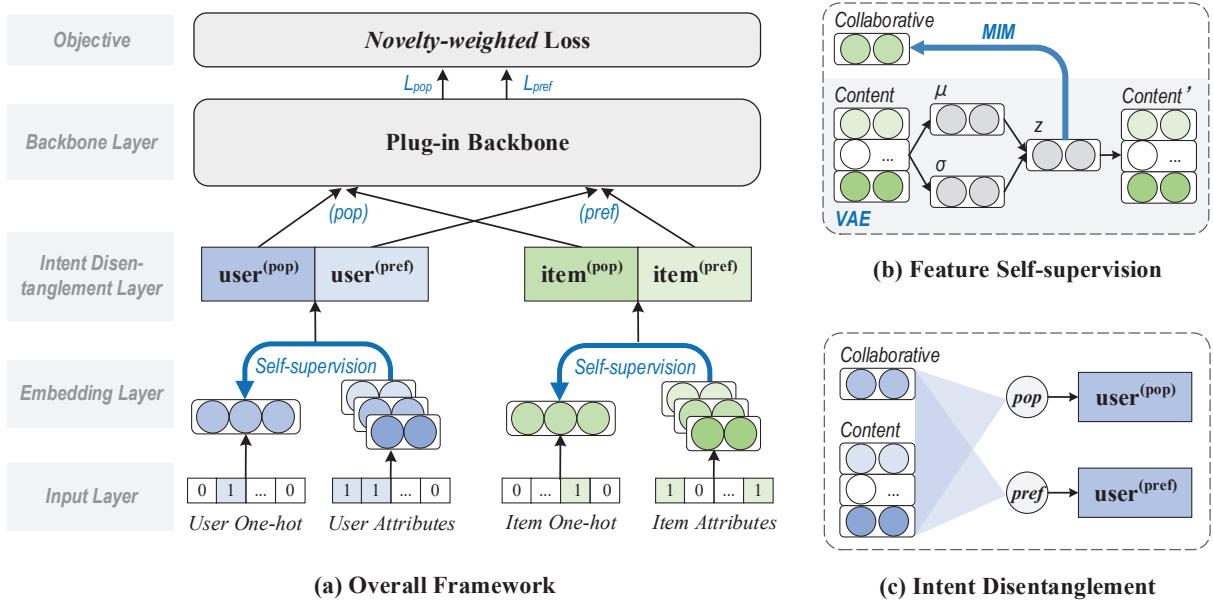}}
		\caption{An overview of our IDS4NR model. (a) Overall framework. (b) Feature self-supervision module for alleviating the cold-start problem. (c) Explicit intent disentanglement module for extracting the specific factors from learned representations.}
		\label{fig-model-overall-1}
	\end{figure*}
	
	Due to the different popularity degree of users and those of items, the users have specific intents to purchase/watch different items. Our target is to find out users' intrinsic intents for different items and eliminate the dominance of popular items, such that we can select the relevant long-tail items and cold-start items for target users to improve the novelty of recommendation.
To this end, we propose an intent disentanglement and feature self-supervision (IDS4NR) model and show its architecture in Figure \ref{fig-model-overall-1}(a).

%Our framework consists of an input and embedding layer, a self-supervision module, an  intent disentanglement layer, and a generic backbone layer.
Our model consists of four layers. We first present an input and embedding layer, which takes users'/items' unique one-hot ID encoding and their corresponding attributes as input and and transform them into low-dimensional latent vectors. We then develop a self-supervision module to map the features from different views to tackle the cold-start problem. Next, we design an intent disentanglement layer to disentangle the latent features into popularity and preference factors. Finally, we introduce a plug-in backbone layer to model the interactions between the user and the item, which can adopt different base models to make the framework generic. %Finally, the loss of popularity and preference intents are weighted by novelty coefficient to control the effect of intent representation to the interaction.

	\subsection{Model Architecture}
	\subsubsection{Input and Embedding Layer}
	\textit{Initial collaborative features.} We set up a lookup table to transform the one-hot representations of each user and item into low-dimensional vector. After transformation, $\bm{p}_u \in \mathbb{R}^D$ and $\bm{q}_v \in \mathbb{R}^D$ are  the latent factor representation of the user $u$ and the item $v$, respectively. We denote them as collaborative features because they are learned from user-item interaction behaviors.
	
	\textit{Initial content features.} Besides the interactions, each user/item is equipped with side information that describes its own characteristics, such as attribute information, text description, and image. To be general, we term them as content features. Specifically, in our scenario, each user or item is associated with a set of attributes from different fields. Below is an example of user attributes:
	\begin{equation} \nonumber
		\underbrace{[0, 1]}_{\textit{gender}} \ \underbrace{[1, 0, 0,... , 0]}_{\textit{age}} \ \underbrace{[0, 1, 0,... , 0]}_{\textit{occupation}}.
		\vspace{-0mm}
	\end{equation}
	Similar to the transformation of collaborative features, we set up an attribute encoding matrix of users and items. The attribute encoding of a user $u$ and an item $v$ is denoted as $[\bm{a}_u^1, \bm{a}_u^2, ..., \bm{a}_u^k]$ and $[\bm{a}_v^1, \bm{a}_v^2, ..., \bm{a}_v^k]$, respectively, where $\bm{a}^i \in \mathbb{R}^D$.
	
	%\subsubsection{Collaborative-Content Self-supervision Module}
	\subsubsection{Feature Self-supervision Module}
	New released items can be a good way to surprise users. However, these cold-start items do not contain any historical interaction data, so the valuable collaborative features cannot be learned from the traditional method. In this paper, since we extend the standard long-tail recommendation to cold-start items, we have to face such \textit{a collaborative feature missing problem}. Below we elaborate our new paradigm on solving this challenge.
	
	The content information of the new item has been proven to be important in solving the cold-start problem. Different from simply fusing the auxiliary content information into the item representation, we believe that it is critical to establish the correlation between the content features and the collaborative features for cold-start items. We term this \emph{the feature self-supervision} between two views of the target item. For example, animation movies are the mainstream entertainment among teenage children. The animation movies (content features) and the behaviors of their audience (collaborative features) have a strong correlation, while the love movies (other content features) and the behaviors of teenage children (collaborative features) have a weak correlation.

In light of this, we first set up the mapping between collaborative and content features, and then generate the missing collaborative features of cold-start items when making recommendations. In the field of representation learning, self-supervised learning leverages input data to construct pre-training tasks and thus can benefit most types of downstream tasks. In our recommendation scenario, we consider the self-supervision as an  auxiliary task, and design a collaborative-content feature self-supervision module.

As shown in Figure \ref{fig-model-overall-1}(b), our  self-supervision module has two key components: \emph{variational autoencoder (VAE)} and \emph{ mutual information maximization (MIM).} Both of them are used to establish the strong correlation between the behaviors of items and their related contents.
Specifically, we adopt VAE to encode content features into the latent space, so that the latent representation has the same distribution as the original one. We further maximize the mutual information  between the generated collaborative features and the inherent content features in order to enhance the representation's discriminative ability.
	
	The VAE structure assumes that data are generated from underlying latent representation. Given the data, the posterior distribution over a set of unobserved variables is approximated by a variational distribution. VAE consists of a generation part and an inference part.
	Take the item $v$ as an example, we denote its content features as the concatenation of all of its attribute encoding, i.e., $\bm{x}_v = \bm{a}_v^1 \oplus ... \oplus \bm{a}_v^k$. In the generation part, the reconstructed content features $\bm{x}'_v$ is generated from its latent variables $\bm{z}_v$ through a generation network like MLP parameterized by $\theta$:
	\begin{equation}\label{equ:vae_generate}
		\bm{x}'_v \sim p_\theta(\bm{x}'_v | \bm{z}_v).
	\end{equation}
	In the inference part, variational inference approximates the true intractable posterior of the latent variable $\bm{z}_v$ by introducing an inference network parameterized by $\phi$	~\cite{VAE_ICLR14}.
	\begin{equation}
		q_\phi (\bm{z}_v) = \mathcal{N}(\bm{\mu}_v, diag(\bm{\sigma}^2_v)),
	\end{equation}
	The objective of variational inference is to optimize the free variational parameters so that the KL-divergence $D_{KL}(q(\bm{z}_v) \| p(\bm{z}_v  \|\bm{x}_v))$ is minimized. With the reparameterization trick, we sample $\bm{\epsilon} \sim N(0,\bm{I})$ and reparameterize $\bm{z}_v =µ_\phi (\bm{x}_v) + \bm{\epsilon} \odot \sigma_{\phi}(\bm{x}_v)$. In this case, the gradient towards $\phi$ can be back-propagated through the sampled $\bm{z}_v$.
	In summary, the loss function in VAE is defined as follows:
	\begin{equation}
		\mathcal{L}_{VAE} = -D_{KL} (q_\phi(\bm{z}_v|\bm{x}_v)\|p(\bm{z}_v)) + \mathbb{E}_{ q_\phi(\bm{z}_v|\bm{x}_v)} [\log p_\theta(\bm{x}_{v}^{'}|\bm{z}_v)]
	\end{equation}
	
Mutual information (MI) is a basic concept in statistics, which measures dependencies between random variables.
%Maximizing mutual information directly is usually intractable. A common practice is to maximize the lower bound of mutual information with an NCE objective ~\cite{NCE_12,SSL_arxiv20}.	
When encoding the the content features into latent space, to increase the discriminative ability, we further identify the correlation between the latent content features and its corresponding collaborative features using the MIM strategy.
	
	Intuitively, the item $v$'s latent content features $\bm{z}_v$ is more relevant to its own collaborative features $\bm{q}_v$, compared with the collaborative features of other items. Therefore, we devise the MIM loss function as:
	\begin{equation}
		\mathcal{L}_{MIM} = -\sum \limits_{(a,v,v')\in \mathcal{O}} \log \sigma(f_D(\bm{z}_v, \bm{q}_v) - f_D(\bm{z}_v, \bm{q}'_v)),
	\end{equation}
	where $(a,v,v')$ are the anchor item attributes, the positive item, and the negative item. We uniformly sample the item $v'$ as the negative one, which does not contain any attributes of $a$ from the entire item space. $f_D({\cdot})$ is the discriminator function that takes two vectors as the input and then scores the agreement between them. We simply implement it as the dot product between two representations to exclude additional parameters ~\cite{MHCN_www21}.
	
	Finally, the whole loss function of the auxiliary feature self-supervision task can be defined as:
	\begin{equation} \label{Equ:SSUP}
		\mathcal{L}_{SS} = \mathcal{L}_{VAE} + \mathcal{L}_{MIM}.
	\end{equation}

	\subsubsection{Intent Disentanglement Module}
Recall that we analyze the users' different levels of conformity in the introduction part, which are ignored by existing novel recommendation methods. In this subsection, we  present our intent disentanglement module to tackle this problem.
%	Once we obtain the meaningful collaborative features and content features, we can perform intent disentanglement in this subsection. Existing studies ~\cite{DisenGCF_sigir20, DisenNews_acl20, MacroRec_nips20, DisenCont_recsys20, DisenRec_kdd20} about disentangled representation in recommendation first define $K$ channels or user intents, and comprehend these latent representation through post-hoc explanation after disentanglement. Different from these methods, we perform intent disentanglement in an explicit fashion and adjust the representation learning of the intents according to items' popularity degree automatically.
	
	To distinguish the reason why users interact popular or niche items, we  set the intents as popularity and preference factors and model the contribution of specific intents to the user-item interactions. In this way, we alleviate the dominance of popularity and increase the novelty naturally. Our proposed intent disentanglement module is shown in Figure \ref{fig-model-overall-1}(c), which has three unique properties.
\begin{itemize}
  \item We present an intent prototype strategy which explicitly disentangles the intents into two   factors (popularity and preference), rather than the parameterized $K$ channels in other disentanglements methods~\cite{DisenGCF_sigir20, DisenNews_acl20, MacroRec_nips20, DisenCont_recsys20, DisenRec_kdd20}.
  \item Both collaborative and content features take effects when disentangling intents, which tallies better with the conformity theory than previous casual model and negative sampling strategy~\cite{Zheng_Disentangling_www21}.
  \item Our disentangled factors are naturally integrated into the end-to-end novel recommendation framework, where the popularity and preference factors account for the accuracy and novelty, respectively.
\end{itemize}

	\textit{Intent prototype strategy.}
We take the user $u$ as an example for illustration. Formally, the user $u$'s overall feature list is denoted as $[\bm{a}_u^0, \bm{a}_u^1, \bm{a}_u^2, ..., \bm{a}_u^k]$, where $\bm{a}_u^0$ is equal to her collaborative feature $\bm{p}_u$ and the rest are $k$ content features.

	We start by defining the prototypical intent representations, i.e., the popularity prototype $\bm{c}^{(pop)} \in \mathbb{R}^D$ and the preference prototype $\bm{c}^{(pref)} \in \mathbb{R}^D$, which are part of model parameters and reflect the position of specific intent in the latent space. We then cluster the initial features according to their distance to two intent prototypes:
	\begin{gather}\label{Equ:cluster}
		p_{u}^{(pop)|i} = \frac{\exp (\bm{c}^{(pop)} \bm{a}_u^i)}{\exp (\bm{c}^{(pop)} \bm{a}_u^i) + \exp (\bm{c}^{(pref)} \bm{a}_u^i)}, \\
		p_{u}^{(pref)|i} = \frac{\exp (\bm{c}^{(pref)} \bm{a}_u^i)}{\exp (\bm{c}^{(pop)} \bm{a}_u^i) + \exp (\bm{c}^{(pref)} \bm{a}_u^i)},
	\end{gather}
	where $i=0,1,2,...,k$. We adopt dot product to measure the similarity between a given input feature vector and an intent prototype.
	
	\textit{Intent aggregation based on  collaborative and content features.}
	The clustering weight $p$ describes the correlation between a specific feature vector and the intent prototype. We now aggregate all features and obtain the integrated representation of the user $u$ under each disentangled intent:
	\begin{gather}\label{Equ:aggregate}
		\bm{p}_{u}^{(pop)} = \text{FFN}(\sum_{i=0}^{k} p_{u}^{(pop)|i} \cdot \bm{a}_{u}^{i}), \\
		\bm{p}_{u}^{(pref)} = \text{FFN}(\sum_{i=0}^{k} p_{u}^{(pref)|i} \cdot \bm{a}_{u}^{i}),
	\end{gather}
	where $\text{FFN}(\bm{x}) = \bm{W}_f \bm{x} + \bm{b}_f$ is a feed-forward network. The item $v$'s popularity and characteristic representation $\bm{q}_{v}^{(pop)}, \bm{q}_{v}^{(pref)}$ can be obtained in the same way. So far, we have disentangled the user/item representation under specific intents. We will later explain how the disentangled intents contribute to the whole framework in the optimization subsection.

	\subsubsection{Backbone Layer}
	After obtaining the user and item representations, we take traditional recommendation methods as a plug-in backbone layer to model the relationship between user and item, which ensures the generality of our framework in enhancing recommendation novelty.
	
	Traditional recommendation methods can be roughly grouped into two types: point-wise and pair-wise~\cite{cdae_wsdm16}. Note that some methods use the list-wise objective, but they are not widely adopted due to the expensive computational cost. %We give a brief introduction to these two types of methods.
	
	\textit{Point-wise objective.}
	The basic idea in point-wise methods is that they only consider the individual user-item pair for ranking prediction. They do not relate other observed user-item pairs to the learning process. This kind of methods takes a triplet $(u,v,y_{uv})$ as the input, where $y_{uv}=1$ indicates that the user $u$ has interacted with the item $v$ (implicit preference), otherwise $y_{uv}=0$. The general form of the point-wise loss function can be defined as:
	\begin{equation}
		\sum \limits_{(u,v,y_{uv})\in \mathcal{O}} \mathcal{L}_{point}^{(intent)}(y_{ui}, \hat{y}_{ui}) + \lambda \varOmega (\theta),
	\end{equation}
	where $intent\in\{pop, pref\}$ and $\varOmega (\theta)$ is a regularization term. A series of traditional methods adopt this kind of optimization strategy and we adopt LFM~\cite{LFM_09} and NCF~\cite{NCF_www17} as representative base models. During this process, we feed the user/item representation under specific intents into the backbone layer, then obtain the predicted score $\hat{y}_{ui}$ and finally calculate the ranking loss according to these base models.
	
	\textit{Pair-wise objective.}
	The pair-wise approaches try to construct the set of item pairs by concerning their relative ordering. It is usually considered to be more suitable for optimizing the top-$N$ recommendation since it compares the relevance of samples instead of getting one ranking score.  This kind of methods inputs a triplet $(u,v^+,v^-)$, where $v^+, v^-$ are the implicitly liked and not observed item by the user $u$. The general form of the pair-wise loss function can be defined as:
	\begin{equation}
		\sum \limits_{(u,v^+,v^-)\in \mathcal{O}} \mathcal{L}_{pair}^{(intent)}(y_{uv^+v^-}, \hat{y}_{uv^+v^-}) + \lambda \varOmega (\theta),
	\end{equation}
	We chose the classic CML~\cite{CML_www17} as a representative base model of this kind of approaches. %and leaving the exploration of more base models in future work.
	
	\subsection{Optimization and Learning}
\subsubsection{Novelty-weighted Recommendation Loss}
	After feeding the disentangled user and item representations into the backbone layer, we can obtain the specific loss $\mathcal{L}^{(intent)}$ ($intent\in\{pop, pref\}$). However, if simply summing the losses without distinction, the learned representation of different intents cannot well reflect the popularity and preference factors. Thus we design a novelty-weighted loss to control the impacts of different intents.
	
	Firstly, inspired by ~\cite{MatchNov_icdm19}, the item novelty can be measured by its popularity degree, namely an item is more likely to be novel when more users have never interacted with it. The novelty score of the item $v$ can be defined as:
	\begin{equation} \label{Equ:novelty_factor}
		\alpha_v' = \log \frac{|U|}{|U_v|},\ \alpha_v = \frac{\alpha_v' - \min( \alpha_{V'})}{\max(\alpha_{V'})-\min(\alpha_{V'})}.
	\end{equation}
	After that, the recommendation loss takes a weighted sum of two intents through the novelty score, where $\sigma(.)$ is sigmoid function for normalization:
	\begin{equation} \label{Equ:loss_rec}
		\mathcal{L}_{Rec} =\left\{
		\begin{array}{l}
			\alpha_v \mathcal{L}^{(pref)} + (1-\alpha_v) \mathcal{L}^{(pop)}, \text{\quad if point-wise,} \\
			\sigma(\alpha_v^+ - \alpha_v^-) \mathcal{L}^{(pref)} + [1 - \sigma(\alpha_v^+ - \alpha_v^-)] \mathcal{L}^{(pop)}, \text{otherwise.}
		\end{array}
		\right.
	\end{equation}
	
	Even if it is a common practice to regard the novelty score as weight, we argue that it is an effective way to align the intents to the popularity and preference factors.

\subsubsection{Training on Disentangled Representations}
Below we discuss the training strategy for point-wise and pair-wise objectives based on the disentangled user and item representations.
(1) For the point-wise learning, the model takes a triplet $(u,v,y_{uv})$ as input. Assuming that the novelty of the item $v$ is high, e.g. $\alpha_v>0.5$, the users' interaction to  $v$ is reasonably due to her preference intent. Therefore, the loss $\mathcal{L}^{(pref)}$ has a larger   weight and it would have more impacts on $\bm{p}_{u}^{(pref)}, \bm{q}_{v}^{(pref)}$. Conversely, the popularity intent becomes the leading factor when the user interacts hot items.
	
	(2) For the pair-wise learning, the model takes a triplet $(u,v^+,v^-)$ as input. Assuming that the novelty of the item $v^+$ is higher than $v^-$, i.e., $\alpha_{v^+}>\alpha_{v^-}$, the user $u$ is more likely to interact with the item $v^+$ for personal preference than $v^-$. Hence $\mathcal{L}^{(pref)}$ has a larger weight, and vice versa.
	
	Finally, the overall loss of our IDS4NR framework contains the loss for the main task of novel recommendation and that for the auxiliary task of feature self-supervision:
	\begin{equation} \label{Equ:joint}
		\mathcal{L} = \mathcal{L}_{Rec} + \gamma \mathcal{L}_{SS},
	\end{equation}
	where $\gamma$ is a constant weighting factor.

\section{Experiments}
	\begin{comment}

	In this section, we conduct extensive experiments on three real-world datasets to validate our proposed IDS4NR model. We aim to answer the following research questions:
	
	\begin{itemize}[leftmargin=*]
		\item \textbf{RQ1:} Does the proposed model outperform the state-of-the-art novelty promoting and cold-start oriented methods in our scenario?
		
		\item \textbf{RQ2:} How do different components such as the intent disentanglement module affect the results of our model?
		
		\item \textbf{RQ3:} How to comprehend the novelty factor in recommender system, and what is the difference between our model and existing methods in recommendation novelty?
		
	\end{itemize}
	\end{comment}
	
	\subsection{Experimental Setup}
	\subsubsection{Datasets}
	
	We use three publicly available datasets from different domains.
	\textbf{MovieLens}\footnote{https://grouplens.org/datasets/movielens/} is a widely adopted dataset in the application domain of recommending movies to users, and we employ the MovieLens-100K version.
	\textbf{Music} and \textbf{Beauty} are chosen from Amazon\footnote{http://jmcauley.ucsd.edu/data/amazon/links.html}. 
%which both contain metadata of diverse products
Following the evaluation settings in ~\cite{A3NCF_ijcai18, Chen_NARRE_www2018}, we take the 5-core version for experiments, where each user or item has at least five interactions.

	In order to be consistent with the implicit feedback setting~\cite{NCF_www17, enmf_tois20}, we transform the detailed rating into a value of 0 or 1, indicating whether a user has rated an item. The statistics of the datasets are shown in Table \ref{tab:dataset}.
	
	\begin{table}[]
		\caption{Statistics of the datasets.}
		\centering
		\footnotesize
		\setlength{\tabcolsep}{2mm}	
		\renewcommand\arraystretch{1.25}
		\label{tab:dataset}
		\begin{tabular}{|c|c|c|c|c|}
			\hline
			Dataset   & \#Users & \#Items & \#Interactions & Sparsity \\ \hline \hline
			MovieLens & 943     & 1682    & 100000         & 93.70\%  \\ \hline
			Music     & 5541    & 3568    & 64706          & 99.67\%  \\ \hline
			Beauty    & 8159    & 5863    & 98566          & 99.79\%  \\ \hline
		\end{tabular}
	\end{table}
	
	\subsubsection{Evaluation Metrics}
	We evaluate the performance of novel recommendation in terms of accuracy, coverage, novelty and the trade-off.
	
	We adopt Recall@N (Rec@N)  ~\cite{CML_www17,Generic_icde18,tailnet_recsys20} as the accuracy metric, which considers whether the ground-truth is ranked amongst the top-$N$ items (normally @5 or @10):
	\begin{equation}
		Rec@N = \frac{1}{|\mathcal{U}|} \sum_{u \in \mathcal{U}} \frac{|\mathcal{I}_u^{Te} \cap \mathcal{P}_u|}{|\mathcal{I}_u^{Te}|},
	\end{equation}
	where $\mathcal{I}_u^{Te}$ is ground-truth list of user $u$ and $\mathcal{P}_u$ is user $u$'s top-$N$ recommendation list.
	
	Coverage@N (Cov@N) ~\cite{YinCLYC_vldb12,Generic_icde18, tailnet_recsys20} is the ratio of the total number of distinct recommended items to the total items:
	\begin{equation}
		Cov@N = \frac{|\cup_{u \in \mathcal{U}} \mathcal{P}_u|}{|\mathcal{I}|}.
	\end{equation}
	
	Following the Pareto principle ~\cite{Pareto_86}, the head items are  the 20\% most popular items, and the rest are long-tail novel items.  We extend the NovAccuracy@N (Nov@N)~\cite{Generic_icde18,MatchNov_icdm19,tailnet_recsys20} to measure how many novel items are in each top-$N$ recommendation list:
	\begin{equation}
		Nov@N = \frac{1}{N|\mathcal{U}|}\sum_{u \in \mathcal{U}} |\mathcal{I}_{Nov} \cap \mathcal{P}_u|,
	\end{equation}
	where $\mathcal{I}_{Nov}$ is novel item set including both long-tail and cold-start items.
	
	Lastly, we employ F-score as a harmonic mean of conflicting accuracy (recall), novelty, and coverage:
	\begin{equation}
		F1@N =  \frac{3 * Rec@N * Nov@N * Cov@N}{Rec@N + Nov@N + Cov@N}.
	\end{equation}

	\subsubsection{Baselines}
	To demonstrate the effectiveness of our proposed IDS4NR model, we compare it with the following state-of-the-art baseline methods. In addition, we adopt different classical base models as backbone layer to show the generality of our framework.
	
	\textit{Long-tail recommendation baselines:}
	\begin{itemize}[leftmargin=*]
		\item  \textbf{GANC}~\cite{Generic_icde18} presents a re-ranking framework to integrate the learned user long-tail preference for accuracy, coverage, and novelty oriented recommendation.
		
		\item \textbf{PPNW}~\cite{MatchNov_icdm19} proposes a loss weighting approach by leveraging users' and items' novelty information for end-to-end novel recommendation.
		
		\item \textbf{TailNet}~\cite{tailnet_recsys20} designs a preference	mechanism which determines users' preference on popular or niche items in session-based long-tail recommendation. We remove its GRU encoder that is irrelevant to our experiments.
		
		\item \textbf{MIRec}~\cite{MIRec_www21} is a dual transfer learning framework which collaboratively transfers knowledge from both model-level and item-level and from head items to tail items.
	\end{itemize}
	
	\textit{Cold-start recommendation baselines:}
	\begin{itemize}[leftmargin=*]
		\item  \textbf{DropoutNet}~\cite{DropoutNet_NeuIPS2017} incorporates content and collaborative information with a neural model. It presents the dropout technique to tackle the cold-start issues.
		
		\item \textbf{HERS}~\cite{HERS_aaai19} employs users’ social relations as user-user graph and builds item-item graph based on the common tags between two items. It aggregates user and item relations for addressing cold-start problem.
		
		\item \textbf{MetaEmb}~\cite{2019-SIGIR-MetaEmb} is a meta-learning approach to cold-start problem. It trains an embedding generator for new IDs through gradient-based meta-learning technique.
		
		\item \textbf{AGNN}~\cite{AGNN_tkde20} develops a graph neural network variant for attribute graph and designs an extended VAE structure for strict cold-start recommendation.
	\end{itemize}
	
	\textit{Classical base models:}
	\begin{itemize}[leftmargin=*]
		\item \textbf{LFM}~\cite{LFM_09} is a classical matrix factorization method for collaborative filtering, which learns the latent factors by alternating least	squares.
		
		\item \textbf{NCF}~\cite{NCF_www17} presents a neural collaborative filtering method that combines multi-layer perceptron with generalized matrix factorization to encode non-linearities.
		
		\item \textbf{CML}~\cite{CML_www17} is a metric learning method, which encodes the user and item into a joint metric space and measures the user-item pair by euclidean distance.
	\end{itemize}

	\subsubsection{Settings}
	To simulate the strict item cold-start scenario, we choose 20\% items which have the latest average interactions chronologically as cold-start items. Note all their interactions are put into the testing set such that they are unseen during training. In addition, to be consistent with normal training scenario, we randomly select 10\% of historical interactions for each user and add them into the testing set, and the remaining data are treated as training set. During the test phrase, our evaluation protocol ranks all unobserved items in the training set for each user~\cite{Generic_icde18,EHCF_aaai20}.
	
	For the baselines, we follow the same hyper-parameter settings if they are reported by the authors and we fine-tune the hyper-parameters if they are not reported. For our IDS4NR, we set the batch size = 128, the initial learning rate = 0.001, the number of max epoch = 40, 40, 30 for the base model LFM, NCF, and CML, respectively. We use Adam~\cite{KingmaB14} as optimizer to self-adapt the learning rate. We sample 4 unobserved items as negative samples for each positive user-item pair~\cite{NCF_www17}. The embedding dimension is set to 50 for both our IDS4NR and all the baselines, to ensure the comparable trainable parameters.
	
	\subsection{Performance Comparison}
		% Please add the following required packages to your document preamble:
	% \usepackage{multirow}
	% \usepackage[normalem]{ulem}
	% \useunder{\uline}{\ul}{}
	\begin{table*}[]
		\caption{The overall performance comparison on three datasets in terms of accuracy, coverage, novelty and trade-off evaluation, respectively. The best performance among all is in bold while the second best one is marked with an underline. }
		\label{tab:performance}
		\centering
		%\small
		\footnotesize
		\setlength{\tabcolsep}{2.6mm}
		\def\arraystretch{1.25}
		\begin{tabular}{|l|l||cc|cc|cc||cc|}
			\hline
			%Dataset & Model & Rec@5 & Rec@10 & Cov@5 & Cov@10 & Nov@5 & Nov@10 & F1@5 & F1@10 \\ \hline \hline
			\multirow{2}{*}{Dataset} & \multicolumn{1}{l||}{\multirow{2}{*}{Model}} & \multicolumn{2}{c|}{Accuracy} & \multicolumn{2}{c|}{Coverage} & \multicolumn{2}{c||}{Novelty} & \multicolumn{2}{c|}{Trade-off} \\ \cline{3-10}
			& \multicolumn{1}{l||}{} & \multicolumn{1}{c|}{Rec@5} & \multicolumn{1}{c|}{Rec@10} & \multicolumn{1}{c|}{Cov@5} & \multicolumn{1}{c|}{Cov@10} & \multicolumn{1}{c|}{Nov@5} & \multicolumn{1}{c||}{Nov@10} & \multicolumn{1}{c|}{F1@5} & \multicolumn{1}{c|}{F1@10} \\ \hline
			
			\multirow{13}{*}{MovieLens} & GANC & 0.0686 & 0.1123 & 0.3565 & 0.4194 & {\ul 0.4445} & {\ul 0.4506} & 0.0375 & 0.0648 \\
			& PPNW & 0.0328 & 0.0562 & 0.4349 & 0.5062 & \textbf{0.6937} & \textbf{0.6896} & 0.0256 & 0.0470 \\
			& TailNet & 0.0761 & 0.1247 & 0.3565 & 0.4497 & 0.2483 & 0.2386 & 0.0297 & 0.0493 \\
			& MIRec & 0.0713 & 0.1260 & 0.3357 & 0.4212 & 0.4627 & 0.4598 & 0.0382 & 0.0727 \\ \cline{2-10}
			& DropoutNet & 0.0811 & 0.1387 & 0.3689 & 0.4385 & 0.2517 & 0.2614 & 0.0322 & 0.0569 \\
			& HERS & 0.0855 & 0.1397 & 0.3701 & 0.4569 & 0.1862 & 0.2170 & 0.0275 & 0.0510 \\
			& MetaEmb & 0.0826 & 0.1296 & 0.4159 & 0.4979 & 0.2430 & 0.2619 & 0.0337 & 0.0570 \\
			& AGNN & \textbf{0.0996} & \textbf{0.1598} & 0.3184 & 0.4147 & 0.1756 & 0.2045 & 0.0281 & 0.0522 \\ \cline{2-10}
			& LFM & {\ul 0.0888} & {\ul 0.1455} & 0.3891 & 0.4699 & 0.1974 & 0.2283 & 0.0303 & 0.0555 \\
			& IDS4NR\scriptsize{(+LFM)} & 0.0870 & 0.1439 & 0.4414 & 0.5721 & 0.3870 & 0.3901 & 0.0487 & 0.0871 \\  \cline{2-10}
			& NCF & 0.0806 & 0.1274 & 0.3951 & 0.4842 & 0.2492 & 0.2745 & 0.0328 & 0.0573 \\
			& IDS4NR\scriptsize{(+NCF)} & 0.0852 & 0.1386 & {\ul 0.4670} & {\ul 0.6155} & 0.3906 & 0.3853 & {\ul 0.0494} & 0.0865 \\ \cline{2-10}
			& CML & 0.0886 & 0.1376 & 0.3755 & 0.4533 & 0.1836 & 0.2121 & 0.0283 & 0.0494 \\
			& IDS4NR\scriptsize{(+CML)} & 0.0752 & 0.1302 & \textbf{0.5519} & \textbf{0.6613} & 0.4296 & 0.4253 & \textbf{0.0506} & \textbf{0.0903} \\ \hline \hline
			
			\multirow{13}{*}{Music} & GANC & 0.0428 & 0.0581 & 0.6304 & 0.6657 & \textbf{0.7515} & \textbf{0.7773} & 0.0427 & 0.0601 \\
			& PPNW & 0.0573 & 0.0911 & 0.4956 & 0.6010 & 0.5426 & 0.5356 & 0.0422 & 0.0717 \\
			& TailNet & 0.0246 & 0.0383 & 0.5850 & 0.6789 & {\ul 0.7324} & {\ul 0.7004} & 0.0236 & 0.0385 \\
			& MIRec & 0.0779 & 0.1297 & 0.4836 & 0.5942 & 0.5547 & 0.5312 & 0.0562 & 0.0979 \\ \cline{2-10}
			& DropoutNet & 0.1010 & 0.1411 & 0.2723 & 0.3600 & 0.1181 & 0.1853 & 0.0198 & 0.0411 \\
			& HERS & \textbf{0.1284} & \textbf{0.1864} & 0.4934 & 0.6351 & 0.2733 & 0.2937 & 0.0580 & 0.0935 \\
			& MetaEmb & 0.0894 & 0.1355 & 0.6990 & 0.7699 & 0.3633 & 0.3701 & 0.0591 & 0.0908 \\
			& AGNN & 0.1121 & {\ul 0.1660} & 0.5707 & 0.6867 & 0.3169 & 0.3439 & 0.0608 & 0.0983 \\ \cline{2-10}
			& LFM & {\ul 0.1147} & 0.1620 & 0.6203 & 0.7223 & 0.2534 & 0.2755 & 0.0547 & 0.0834 \\
			& IDS4NR\scriptsize{(+LFM)} & 0.0891 & 0.1366 & 0.6853 & 0.8324 & 0.5006 & 0.4983 & 0.0719 & 0.1158 \\ \cline{2-10}
			& NCF & 0.0982 & 0.1419 & 0.6231 & 0.7321 & 0.2785 & 0.3073 & 0.0511 & 0.0810 \\
			& IDS4NR\scriptsize{(+NCF)} & 0.0906 & 0.1373 & \textbf{0.9030} & \textbf{0.9747} & 0.5349 & 0.5243 & \textbf{0.0859} & \textbf{0.1287} \\ \cline{2-10}
			& CML & 0.0856 & 0.1257 & 0.5923 & 0.7433 & 0.1916 & 0.2093 & 0.0335 & 0.0544 \\
			& IDS4NR\scriptsize{(+CML)} & 0.1123 & 0.1643 & {\ul 0.7363} & {\ul 0.8156} & 0.4331 & 0.4330 & {\ul 0.0838} & {\ul 0.1232} \\ \hline \hline
			
			\multirow{13}{*}{Beauty} & GANC & 0.0041 & 0.0081 & 0.4432 & 0.4862 & \textbf{0.8850} & \textbf{0.8942} & 0.0036 & 0.0076 \\
			& PPNW & 0.0117 & 0.0189 & 0.4374 & 0.5415 & 0.5952 & 0.5920 & 0.0088 & 0.0158 \\
			& TailNet & 0.0060 & 0.0089 & 0.5336 & 0.6358 & {\ul 0.7290} & {\ul 0.6713} & 0.0055 & 0.0086 \\
			& MIRec & 0.0220 & 0.0370 & 0.4847 & 0.5954 & 0.5729 & 0.5533 & 0.0170 & 0.0309 \\ \cline{2-10}
			& DropoutNet & {\ul 0.0338} & {\ul 0.0510} & 0.1930 & 0.2754 & 0.0691 & 0.1214 & 0.0045 & 0.0114 \\
			& HERS & \textbf{0.0432} & \textbf{0.0650} & 0.4680 & 0.6232 & 0.2375 & 0.2690 & 0.0192 & 0.0341 \\
			& MetaEmb & 0.0272 & 0.0450 & 0.6929 & 0.7676 & 0.3885 & 0.3967 & 0.0198 & 0.0340 \\
			& AGNN & 0.0263 & 0.0403 & 0.5377 & 0.6518 & 0.3124 & 0.3388 & 0.0151 & 0.0259 \\ \cline{2-10}
			& LFM & 0.0242 & 0.0401 & 0.5558 & 0.6906 & 0.1978 & 0.2278 & 0.0102 & 0.0197 \\
			& IDS4NR\scriptsize{(+LFM)} & 0.0270 & 0.0426 & 0.6029 & 0.7115 & 0.4006 & 0.4128 & 0.0190 & 0.0321 \\ \cline{2-10}
			& NCF & 0.0201 & 0.0317 & 0.6286 & 0.7264 & 0.3037 & 0.3319 & 0.0121 & 0.0210 \\
			& IDS4NR\scriptsize{(+NCF)} & 0.0254 & 0.0413 & \textbf{0.9026} & \textbf{0.9737} & 0.5387 & 0.5363 & {\ul 0.0253} & {\ul 0.0417} \\ \cline{2-10}
			& CML & 0.0211 & 0.0314 & 0.6510 & 0.7963 & 0.2204 & 0.2396 & 0.0102 & 0.0168 \\
			& IDS4NR\scriptsize{(+CML)} & 0.0293 & 0.0477 & {\ul 0.7919} & {\ul 0.8608} & 0.4958 & 0.4905 & \textbf{0.0262} & \textbf{0.0432} \\ \hline
		\end{tabular}
	\end{table*}
	
	The performance comparison between our proposed IDS4NR model and the baselines on three datasets is reported in Table \ref{tab:performance}. From the results, we have the following important observations.
	
	Firstly, it is clear that our IDS4NR significantly outperforms all the baselines in terms of the trade-off F1 of accuracy, coverage, and novelty on three datasets. The relative improvement of our model over the best baselines on three datasets are 32.46\%, 45.35\%, 32.32\% on F1@5, and 24.21\%, 31.46\%, 26.69\% on F1@10, respectively. It verifies the superiority of our proposed framework with the novelty-regulated intent disentanglement and collaborative-content feature self-supervision module.
	
	Secondly, compared with four long-tail recommendation baselines,  our IDS4NR achieves the superior performance in terms of accuracy and coverage. Although IDS4NR cannot get enhancement on the novelty metric over these baselines, it outperforms its base model by a large margin in terms of novelty. Among four baselines, GANC, PPNW, and TailNet emphasize the impact of niche items through the re-ranking strategy, or the modified loss function and adjusted user representation, and they do not consider the inherent mechanism that drives users to interact with the items. Consequently, they achieve high novelty but sacrifice the coverage of entire item space. Meanwhile, MIRec transfers the knowledge from head items to tail ones and is better than three other baselines, but it is still inferior to our model.

	\begin{table*}[!htb]
		\caption{Results for ablation study (backbone - \textit{CML})}
		\label{tab:ablation-cml}
		\centering
		%\small
		\footnotesize
		\setlength{\tabcolsep}{2.2mm}
		\def\arraystretch{1.25}
		\begin{tabular}{|l|l||cc|cc|cc||cc|}
			\hline
			\multirow{2}{*}{Dataset} & \multicolumn{1}{l||}{\multirow{2}{*}{Model}} & \multicolumn{2}{c|}{Accuracy} & \multicolumn{2}{c|}{Coverage} & \multicolumn{2}{c||}{Novelty} & \multicolumn{2}{c|}{Trade-off} \\ \cline{3-10}
			& \multicolumn{1}{l||}{} & \multicolumn{1}{c|}{Rec@5} & \multicolumn{1}{c|}{Rec@10} & \multicolumn{1}{c|}{Cov@5} & \multicolumn{1}{c|}{Cov@10} & \multicolumn{1}{c|}{Nov@5} & \multicolumn{1}{c||}{Nov@10} & \multicolumn{1}{c|}{F1@5} & \multicolumn{1}{c|}{F1@10} \\ \hline \hline
			\multirow{4}{*}{MovieLens} & IDS4NR & 0.0752 & 0.1302 & \textbf{0.5519} & \textbf{0.6613} & \textbf{0.4296} & \textbf{0.4253} & \textbf{0.0506} & \textbf{0.0903} \\
			& IDS4NR$_\text{w/o-SS}$ & 0.0793 & 0.1298 & 0.4295 & 0.5436 & 0.4025 & 0.4237 & 0.0451 & 0.0817 \\
			& IDS4NR$_\text{w/o-SS, Exp}$ & 0.0767 & 0.1303 & 0.4238 & 0.5303 & 0.3687 & 0.3825 & 0.0413 & 0.0760 \\
			& IDS4NR$_\text{w/o-SS, ID}$ & \textbf{0.0966} & \textbf{0.1576} & 0.3196 & 0.4260 & 0.2803 & 0.3101 & 0.0373 & 0.0699 \\ \hline \hline
			
			\multirow{4}{*}{Music} & IDS4NR & 0.1123 & 0.1643 & 0.7363 & 0.8156 & \textbf{0.4331} & \textbf{0.4330} & \textbf{0.0838} & \textbf{0.1232} \\
			& IDS4NR$_\text{w/o-SS}$ & 0.1099 & 0.1591 & \textbf{0.7615} & \textbf{0.8814} & 0.4151 & 0.4350 & 0.0810 & 0.1230 \\
			& IDS4NR$_\text{w/o-SS, Exp}$ & 0.1087 & 0.1582 & 0.7329 & 0.8548 & 0.4018 & 0.4212 & 0.0772 & 0.1192 \\
			& IDS4NR$_\text{w/o-SS, ID}$ & \textbf{0.1521} & \textbf{0.2198} & 0.5362 & 0.6791 & 0.2406 & 0.2768 & 0.0633 & 0.1054 \\ \hline \hline
			
			\multirow{4}{*}{Beauty} & IDS4NR & 0.0293 & 0.0477 & 0.7919 & 0.8608 & \textbf{0.4958} & \textbf{0.4905} & \textbf{0.0262} & \textbf{0.0432} \\
			& IDS4NR$_\text{w/o-SS}$ & 0.0260 & 0.0432 & \textbf{0.8040} & \textbf{0.9082} & 0.4703 & 0.4867 & 0.0227 & 0.0398 \\
			& IDS4NR$_\text{w/o-SS, Exp}$ & 0.0254 & 0.0410 & 0.7879 & 0.8995 & 0.4641 & 0.4804 & 0.0218 & 0.0374 \\
			& IDS4NR$_\text{w/o-SS, ID}$ & \textbf{0.0478} & \textbf{0.0703} & 0.4600 & 0.6336 & 0.2238 & 0.2781 & 0.0201 & 0.0378 \\ \hline
		\end{tabular}
	\end{table*}

	Thirdly, compared with four cold-start recommendation baselines, our IDS4NR gets remarkable improvements in terms of coverage and novelty while keeping a comparable accuracy performance. These baselines methods either learn a better representation for cold-start items by incorporating the neighbor information in the form of graph neural network (HERS and AGNN), or simulate and adapt to the cold-start scenario by dropout technique (DropoutNet) or meta-learning (MetaEmb), and thus accurately capture the user preference for cold-start items. However, they ignore the long-tail problem under normal circumstances and do not perform well  in many cases of novel recommendation.
	
	Finally, for the base model LFM, NCF, and CML, these traditional methods do not take into account the novelty factor, resulting in high accuracy but low coverage and novelty. However, when we integrate them as the backbone layer in our model, we can get dramatic trade-off F1 increases. For example, the relative improvement of F1@5 are 60.73\%, 50.61\%, and 78.80\% on MovieLens, 31.44\%, 68.10\%, and 150.15\% on Music, and 86.27\%, 109.09\%, and 156.86\% on Beauty. These results prove that our IDS4NR can enhance the overall performance for various types of classical methods.

\subsection{Detailed Study on IDS4NR}
	\subsubsection{Ablation Study}
Given the appealing performance of IDS4NR, we further examine how the individual components in our model contribute to the improved performance on novel recommendation. Concretely, we conduct a series of ablation studies to investigate the effects of explicit intent disentanglement and collaborative-content self-supervision:
	
	\begin{itemize}
		\item IDS4NR$_\text{w/o-SS}$: This variant removes the feature self-supervision module from the IDS4NR framework.
		\item IDS4NR$_\text{w/o-SS, Exp}$: Our IDS4NR uses an item novelty score $\alpha_v$  to control the impacts of disentangled representation of popularity and preference intents. This variant is based on IDS4NR$_\text{w/o-SS}$ and further replaces $\alpha_v$ in Eq. \ref{Equ:loss_rec} with 0.5 for treating all samples equally.
		\item IDS4NR$_\text{w/o-SS, ID}$: This variant removes the intent disentanglement module completely. To ensure utilizing the content information as the standard IDS4NR, we take the average of collaborative feature and all kinds of attributes of users/items as the input of backbone layer on the basis of IDS4NR$_\text{w/o-SS}$. 
	\end{itemize}
	
For clarity, we choose the CML backbone as the representative to perform the ablation and other parameter studies. 
We first present The ablation results  on three datasets are shown in Table~\ref{tab:ablation-cml}.
%The ablation results of backbone LFM and CML on three datasets are shown in Table \ref{tab:ablation-lfm} and Table \ref{tab:ablation-cml}, resepectively.
	
	Firstly, we investigate the impacts of feature self-supervision module. We find that the overall trade-off performance for IDS4NR$_\text{w/o-SS}$ decline on all three datasets compared with the standard IDS4NR. This indicates that the feature self-supervision component captures the relationship between the collaborative feature and content feature and well models the user preference for cold-start items, thus it is an essential part of IDS4NR.
	
	Next, we examine the impacts of explicit intent disentanglement module. It is clear that both the coverage and novelty of IDS4NR$_\text{w/o-SS, Exp}$ decrease a lot if compared with IDS4NR$_\text{w/o-SS}$. The reason might be that the intent disentanglement without any constraint cannot learn the specific factors explicitly and will limit the  ability of learning representations for  users' intents on popular/niche items.
	
Finally, the overall performance of the variant IDS4NR$_\text{w/o-SS, ID}$ decreases dramatically along with a few increased accuracy. After removing the intent disentanglement, the model degrades into a naive content-based method, and focuses on the items having similar contents with users' historical interactions. With the limited recommendation scope, the accuracy of the model may increase a bit but the coverage and novelty performance become extremely poor.

\subsubsection{Parameter Study}
\begin{comment}	
	\begin{figure*}[]
		\centerline{\includegraphics[width=0.8\textwidth]{dimension_lfm.eps}}
		\caption{Impacts of latent vector dimension $D$ (backbone - \textit{LFM})}
		\label{dimension_lfm}
	\end{figure*}
\end{comment}	
	\begin{figure*}[]
		\centerline{\includegraphics[width=0.8\textwidth]{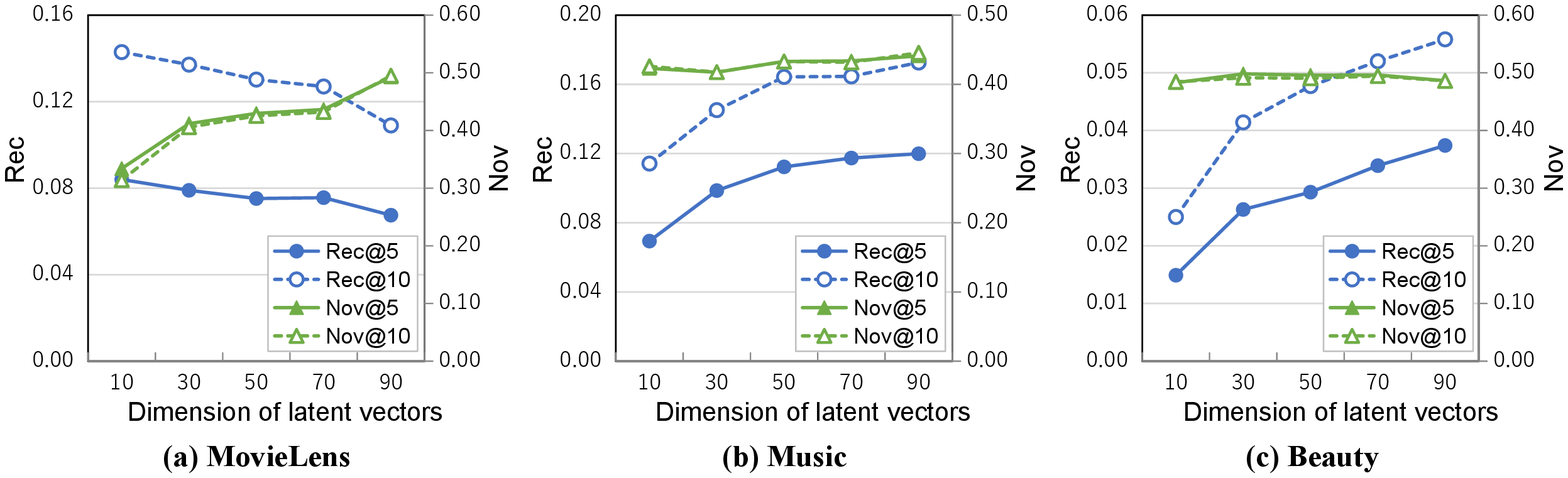}}
		\caption{Impacts of latent vector dimension $D$ (backbone - \textit{CML})}
		\label{dimension_cml}
	\end{figure*}

\begin{comment}	
	\begin{figure*}[]
		\centerline{\includegraphics[width=0.8\textwidth]{weight_lfm.eps}}
		\caption{Impacts of weighting factor $\gamma$ (backbone - \textit{LFM})}
		\label{weight_lfm}
	\end{figure*}
\end{comment}	
	\begin{figure*}[]
		\centerline{\includegraphics[width=0.8\textwidth]{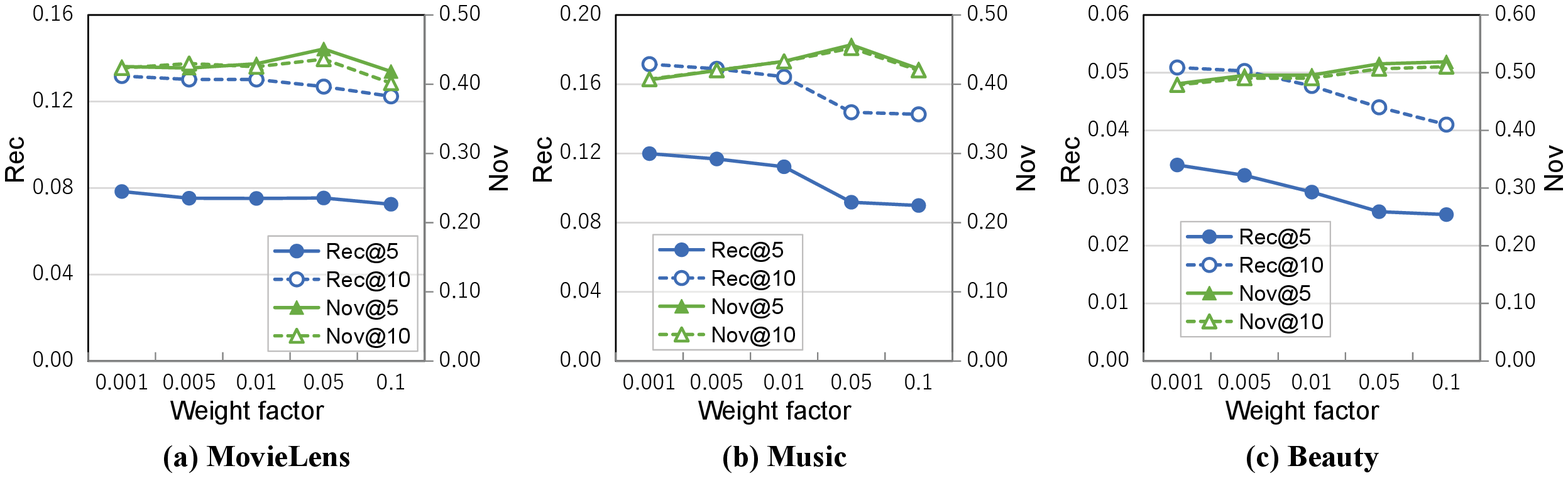}}
		\caption{Impacts of weighting factor $\gamma$ (backbone - \textit{CML})}
		\label{weight_cml}
	\end{figure*}
	
We investigate the impacts of hyper-parameters in IDS4NR, including  the dimension of the latent vector $D$ and the weighting factor $\gamma$ for feature self-supervision loss. We first fix $\gamma$ to 0.01 and vary $D$, and then  fix $D$ to 50 and vary $\gamma$. 
The results of IDS4NR with CML as backbone layer for the varying parameter $D$ and $\gamma$  are shown in  Figure \ref{dimension_cml}  and Figure \ref{weight_cml}, respectively.

%Due to the space limitation, we only present the result of IDS4NR taking LFM and CML as backbone layer of accuracy and novelty performance. Figure \ref{dimension_lfm} and Figure \ref{dimension_cml} show the results by varying the dimensionality in the set of \{10, 30, 50, 70, 90\} of backbone LFM and CML, respectively. Figure \ref{weight_lfm} and Figure \ref{weight_cml} show the results by tuning the weighting factor amongst \{0.001, 0.005, 0.01, 0.05, 0.1\} of backbone LFM and CML, respectively.

We can see from Figure \ref{dimension_cml} that accuracy and novelty show opposite trends in all datasets. This is consistent with the problem setting and the results in previous studies. 
%When accuracy increases, novelty decreases gradually, and vice versa. 
Movielens is a small dataset with relatively dense user-item interactions. The model can already learn a good feature representation even with a small dimensionality ($D$=10). %As it increases gradually, the accuracy decreases. 
In contrast, Music and Beauty are even sparser. The larger dimension improves the representation ability of features with more latent factors of users and items, and thus the model reaches a better accuracy, but its novelty declines at the same time. Hence we set $D$ as 50 for a better balance in most cases.
	
Our IDS4NR takes feature self-supervision learning as an auxiliary task to enhance the collaborative features of cold-start items, and the weighting factor $\gamma$ controls the proportion of self-supervision task. As we can seen in Figure~\ref{weight_cml}, with the increase of $\gamma$, the accuracy on different datasets gradually decreases and the novelty increases, and this trend is more obvious when $\gamma$ exceeds 0.01.
When $\gamma$ is small, it can improve the performance for cold start items as an auxiliary task and thus enhances the recommendation novelty. However, when $\gamma$ is too large, the self-supervision task dominates the loss and interfere the learning of the main task, leading to a sharp decline of recommendation accuracy. %In order to achieve a balance, we set it to 0.01.

\subsubsection{Analysis on Computational Cost}
\begin{comment}	
	\begin{figure*}[]
		\centerline{\includegraphics[width=0.8\textwidth]{loss_lfm.eps}}
		\caption{Training curves of the overall loss. (backbone - \textit{LFM})}
		\label{loss_lfm}
	\end{figure*}
\end{comment}
	
	\begin{figure*}[]
		\centerline{\includegraphics[width=0.8\textwidth]{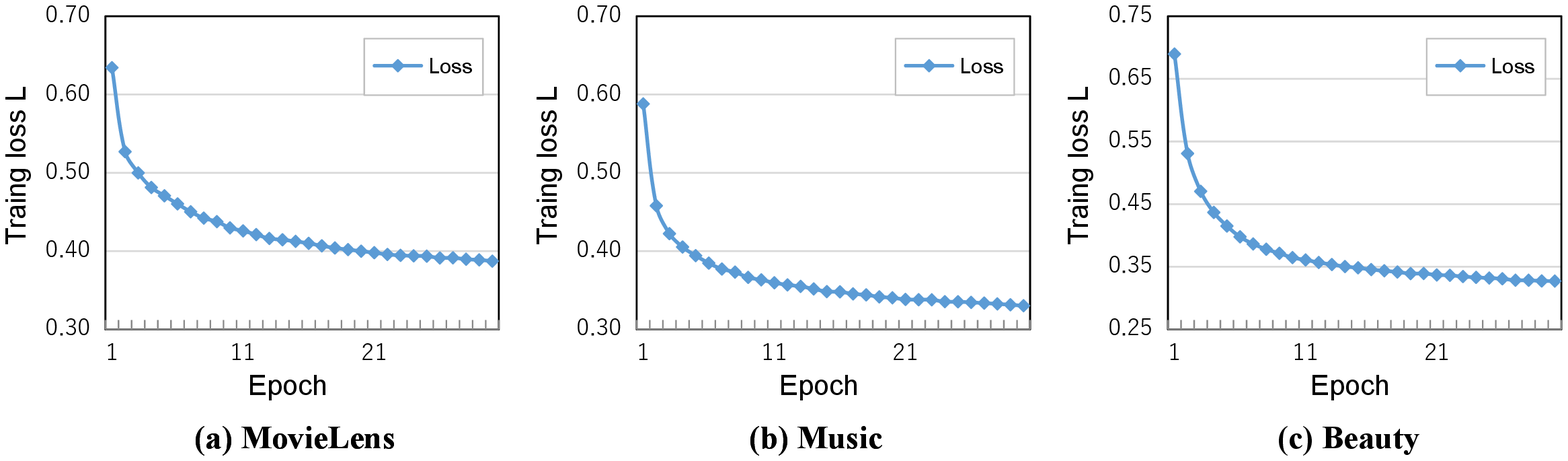}}
		\caption{Training curves of the overall loss. (backbone - \textit{CML})}
		\label{loss_cml}
	\end{figure*}
		
	To investigate the training process of model, we plot the model's training curves in Figure \ref{loss_cml}.  As can be seen, the loss declines quickly first and gradually slows down. It finally reaches the convergence when the training epoch increases, which proves that our model is stable and easy to train.

	We further compare the model's space complexity by presenting the trainable parameter number of our model and that of base models in Table \ref{tab:param}.
	The parameter number of our IDS4NR and its different base models mainly depends on the user/item embedding matrix. Compared to the base models, our framework introduces  additional user/item attributes, so the increased cost lies primarily in the attribute embedding. For the intent disentanglement and self-supervision module in the framework, they need a few weight matrix at a lower cost. Note that NCF has the largest parameter scale on Music and Beauty because it uses the independent embedding layer in its GMF and MLP module, which is equivalent to expanding the dimension of embedding vectors. When we take it as the backbone layer, these two modules are shared by the disentangled user/item representation and thus decrease the parameter number. %and no additional parameters are introduced.
	\begin{table}[!htb]
		\caption{Parameter number of models on three datasets.}
		\renewcommand{\arraystretch}{1.1}
		\setlength{\tabcolsep}{2mm}
		\centering
		\footnotesize
		\label{tab:param}
		\begin{tabular}{|l|ccc|}
			\hline
			\multirow{2}{*}{Model} & \multicolumn{3}{|c|}{Model size ($\times 10^6$)} \\ \cline{2-4}
			& \multicolumn{1}{|l}{MovieLens} & \multicolumn{1}{l}{Music} & \multicolumn{1}{l|}{Beauty} \\ \hline \hline
			LFM & 0.13 & 0.46 & 0.72 \\
			IDS4NR\scriptsize{(+LFM)} & 0.57 & 0.78 & 1.19 \\ \hline
			NCF & 0.28 & 0.93 & 1.42 \\
			IDS4NR\scriptsize{(+NCF)} & 0.60 & 0.84 & 1.24 \\ \hline
			CML & 0.13 & 0.46 & 0.70 \\
			IDS4NR\scriptsize{(+CML)} & 0.52 & 0.76 & 1.16 \\ \hline
		\end{tabular}
	\end{table}

\section{Conclusion}
In this paper, we propose a new model IDS4NR for novel recommendation. We distinguish our model with existing novelty-oriented methods in two key issues. Firstly, we realize that the varying degrees of users' intent to popular and niche items are the inherent factors in determining the trade-off between accuracy and novelty in novel recommendation. We then implement this idea with the intent disentanglement integrated end-to-end framework. Secondly, we extend the definition of standard long-tail recommendation  to cover the cold-start items. We also propose a self-supervision strategy to solve the missing collaborative feature problem which is a big obstacle when recommending cold-start items. 
%In addition, our model provides a general framework which enables various base models as the plug-in backbone.
To validate our thoughts, we conduct extensive comparison experiments and deep analysis studies on three real-world datasets. The results prove that our proposed IDS4NR model achieves a new state-of-the-art trade-off performance on the novel recommendation task for both the standard long-tail items and the extended cold-start items.
	
	\begin{comment}
	% use section* for acknowledgment
	\ifCLASSOPTIONcompsoc
	% The Computer Society usually uses the plural form
	\section*{Acknowledgments}
	\else
	% regular IEEE prefers the singular form
	\section*{Acknowledgment}
	\fi
	
	The work described in this paper was supported in part by the NSFC project (61572376, 91646206).
	\end{comment}
	
	% Can use something like this to put references on a page
	% by themselves when using endfloat and the captionsoff option.
	\ifCLASSOPTIONcaptionsoff
	\newpage
	\fi
	
	\vspace{-2mm}
	\bibliography{ss4nr}

% Generated by IEEEtran.bst, version: 1.13 (2008/09/30)
\begin{thebibliography}{10}
\providecommand{\url}[1]{#1}
\csname url@samestyle\endcsname
\providecommand{\newblock}{\relax}
\providecommand{\bibinfo}[2]{#2}
\providecommand{\BIBentrySTDinterwordspacing}{\spaceskip=0pt\relax}
\providecommand{\BIBentryALTinterwordstretchfactor}{4}
\providecommand{\BIBentryALTinterwordspacing}{\spaceskip=\fontdimen2\font plus
\BIBentryALTinterwordstretchfactor\fontdimen3\font minus
  \fontdimen4\font\relax}
\providecommand{\BIBforeignlanguage}[2]{{%
\expandafter\ifx\csname l@#1\endcsname\relax
\typeout{** WARNING: IEEEtran.bst: No hyphenation pattern has been}%
\typeout{** loaded for the language `#1'. Using the pattern for}%
\typeout{** the default language instead.}%
\else
\language=\csname l@#1\endcsname
\fi
#2}}
\providecommand{\BIBdecl}{\relax}
\BIBdecl

\bibitem{AdomaviciusK_tkde12}
G.~Adomavicius and Y.~Kwon, ``Improving aggregate recommendation diversity
  using ranking-based techniques,'' \emph{{IEEE} Trans. Knowl. Data Eng.},
  vol.~24, no.~5, pp. 896--911, 2012.

\bibitem{YinCLYC_vldb12}
H.~Yin, B.~Cui, J.~Li, J.~Yao, and C.~Chen, ``Challenging the long tail
  recommendation,'' \emph{Proc. {VLDB} Endow.}, vol.~5, no.~9, pp. 896--907,
  2012.

\bibitem{Park_tkde13}
Y.~Park, ``The adaptive clustering method for the long tail problem of
  recommender systems,'' \emph{{IEEE} Trans. Knowl. Data Eng.}, vol.~25, no.~8,
  pp. 1904--1915, 2013.

\bibitem{Oestreicher-SingerS_misq12}
G.~Oestreicher{-}Singer and A.~Sundararajan, ``Recommendation networks and the
  long tail of electronic commerce,'' \emph{{MIS} Q.}, vol.~36, no.~1, pp.
  65--83, 2012.

\bibitem{Generic_icde18}
Z.~Zolaktaf, R.~Babanezhad, and R.~Pottinger, ``A generic top-n recommendation
  framework for trading-off accuracy, novelty, and coverage,'' in \emph{ICDE},
  2018, pp. 149--160.

\bibitem{MatchNov_icdm19}
K.~Lo and T.~Ishigaki, ``Matching novelty while training: Novel recommendation
  based on personalized pairwise loss weighting,'' in \emph{ICDM}, 2019, pp.
  468--477.

\bibitem{tailnet_recsys20}
S.~Liu and Y.~Zheng, ``Long-tail session-based recommendation,'' in
  \emph{RecSys}, 2020, pp. 509--514.

\bibitem{MIRec_www21}
Y.~Zhang, D.~Z. Cheng, T.~Yao, X.~Yi, L.~Hong, and E.~H. Chi, ``A model of two
  tales: Dual transfer learning framework for improved long-tail item
  recommendation,'' \emph{CoRR}, vol. abs/2010.15982, 2020.

\bibitem{Lascu_1999}
D.-N. Lascu and G.~Zinkhan, ``Consumer conformity: Review and applications for
  marketing theory and practice,'' \emph{Journal of Marketing Theory and
  Practice}, vol.~7, no.~3, pp. 1--12, 1999.

\bibitem{Park_2010}
J.~Park and R.~Feinberg, ``E‐formity: consumer conformity behaviour in
  virtual communities,'' \emph{Journal of Research in Interactive Marketing},
  vol.~4, no.~3, pp. 197--213, 2010.

\bibitem{ilike_recsys15}
K.~Kapoor, V.~Kumar, L.~G. Terveen, J.~A. Konstan, and P.~R. Schrater, ``"i
  like to explore sometimes": Adapting to dynamic user novelty preferences,''
  in \emph{RecSys}, 2015, pp. 19--26.

\bibitem{GraphSAGE_NeuIPS17}
W.~Hamilton, Z.~Ying, and J.~Leskovec, ``Inductive representation learning on
  large graphs,'' in \emph{NeuIPS}, 2017, pp. 1024--1034.

\bibitem{IGMC_iclr20}
M.~Zhang and Y.~Chen, ``Inductive matrix completion based on graph neural
  networks,'' in \emph{ICLR}, 2020.

\bibitem{2020-TKDE-FMFC}
Y.~Zhu, J.~Lin, S.~He, B.~Wang, Z.~Guan, H.~Liu, and D.~Cai, ``Addressing the
  item cold-start problem by attribute-driven active learning,'' \emph{{IEEE}
  Trans. Knowl. Data Eng.}, vol.~32, no.~4, pp. 631--644, 2020.

\bibitem{Aharon-Active-Recsys15}
M.~Aharon, O.~Anava, N.~Avigdor{-}Elgrabli, D.~Drachsler{-}Cohen, S.~Golan, and
  O.~Somekh, ``Excuseme: Asking users to help in item cold-start
  recommendations,'' in \emph{RecSys}, 2015, pp. 83--90.

\bibitem{StarGCN_ijcai19}
J.~Zhang, X.~Shi, S.~Zhao, and I.~King, ``Star-gcn: Stacked and reconstructed
  graph convolutional networks for recommender systems,'' in \emph{IJCAI},
  2019, pp. 4264--4270.

\bibitem{HERS_aaai19}
L.~Hu, S.~Jian, L.~Cao, Z.~Gu, Q.~Chen, and A.~Amirbekyan, ``{HERS:} modeling
  influential contexts with heterogeneous relations for sparse and cold-start
  recommendation,'' in \emph{AAAI}, 2019, pp. 3830--3837.

\bibitem{AGNN_tkde20}
T.~Qian, Y.~Liang, Q.~Li, and H.~Xiong, ``Attribute graph neural networks for
  strict cold start recommendation,'' \emph{IEEE Transactions on Knowledge and
  Data Engineering}, pp. 1--1, 2020.

\bibitem{TrustSVD_aaai15}
G.~Guo, J.~Zhang, and N.~Yorke-Smith, ``Trustsvd: Collaborative filtering with
  both the explicit and implicit influence of user trust and of item ratings,''
  in \emph{AAAI}, 2015, pp. 123--125.

\bibitem{NFM_sigir17}
X.~He and T.-S. Chua, ``Neural factorization machines for sparse predictive
  analytics,'' in \emph{SIGIR}, 2017, pp. 355--364.

\bibitem{RelationalCF_sigir19}
X.~Xin, X.~He, Y.~Zhang, Y.~Zhang, and J.~Jose, ``Relational collaborative
  filtering: Modeling multiple item relations for recommendation,'' in
  \emph{SIGIR}, 2019, pp. 125--134.

\bibitem{2017-NeuIPS-LWA}
M.~Vartak, A.~Thiagarajan, C.~Miranda, J.~Bratman, and H.~Larochelle, ``A
  meta-learning perspective on cold-start recommendations for items,'' in
  \emph{NeuIPS}, 2017, pp. 6904--6914.

\bibitem{2019-SIGIR-MetaEmb}
F.~Pan, S.~Li, X.~Ao, P.~Tang, and Q.~He, ``Warm up cold-start advertisements:
  Improving {CTR} predictions via learning to learn {ID} embeddings,'' in
  \emph{SIGIR}, 2019, pp. 695--704.

\bibitem{DropoutNet_NeuIPS2017}
M.~Volkovs, G.~Yu, and T.~Poutanen, ``Dropoutnet: Addressing cold start in
  recommender systems,'' in \emph{NeuIPS}, 2017, pp. 4957--4966.

\bibitem{betaVAE_iclr17}
I.~Higgins, L.~Matthey, A.~Pal, C.~Burgess, X.~Glorot, M.~Botvinick,
  S.~Mohamed, and A.~Lerchner, ``beta-vae: Learning basic visual concepts with
  a constrained variational framework,'' in \emph{ICLR}.\hskip 1em plus 0.5em
  minus 0.4em\relax OpenReview.net, 2017.

\bibitem{HamaguchiSN_cvpr19}
R.~Hamaguchi, K.~Sakurada, and R.~Nakamura, ``Rare event detection using
  disentangled representation learning,'' in \emph{CVPR}, 2019, pp. 9327--9335.

\bibitem{MacroRec_nips20}
J.~Ma, C.~Zhou, P.~Cui, H.~Yang, and W.~Zhu, ``Learning disentangled
  representations for recommendation,'' in \emph{NIPS}, 2019, pp. 5712--5723.

\bibitem{DisenCont_recsys20}
Y.~Zhang, Z.~Zhu, Y.~He, and J.~Caverlee, ``Content-collaborative
  disentanglement representation learning for enhanced recommendation,'' in
  \emph{RecSys}.\hskip 1em plus 0.5em minus 0.4em\relax {ACM}, 2020, pp.
  43--52.

\bibitem{DisenNews_acl20}
L.~Hu, S.~Xu, C.~Li, C.~Yang, C.~Shi, N.~Duan, X.~Xie, and M.~Zhou, ``Graph
  neural news recommendation with unsupervised preference disentanglement,'' in
  \emph{ACL}.\hskip 1em plus 0.5em minus 0.4em\relax Association for
  Computational Linguistics, 2020, pp. 4255--4264.

\bibitem{DisenGCF_sigir20}
X.~Wang, H.~Jin, A.~Zhang, X.~He, T.~Xu, and T.~Chua, ``Disentangled graph
  collaborative filtering,'' in \emph{SIGIR}.\hskip 1em plus 0.5em minus
  0.4em\relax {ACM}, 2020, pp. 1001--1010.

\bibitem{DisenRec_kdd20}
J.~Ma, C.~Zhou, H.~Yang, P.~Cui, X.~Wang, and W.~Zhu, ``Disentangled
  self-supervision in sequential recommenders,'' in \emph{KDD}.\hskip 1em plus
  0.5em minus 0.4em\relax {ACM}, 2020, pp. 483--491.

\bibitem{Zheng_Disentangling_www21}
Y.~Zheng, C.~Gao, X.~Li, X.~He, Y.~Li, and D.~Jin, ``Disentangling user
  interest and conformity for recommendation with causal embedding,'' in
  \emph{WWW}, 2021.

\bibitem{Venkatesan_1966}
M.~Venkatesan, ``"experimental study of consumer behavior conformity and
  independence,'' \emph{Journal of Marketing Research}, vol.~3, pp. 384--387,
  1966.

\bibitem{SSL_arxiv20}
X.~Liu, F.~Zhang, Z.~Hou, Z.~Wang, L.~Mian, J.~Zhang, and J.~Tang,
  ``Self-supervised learning: Generative or contrastive,'' \emph{CoRR}, vol.
  abs/2006.08218, 2020.

\bibitem{dae_icml08}
P.~Vincent, H.~Larochelle, Y.~Bengio, and P.~Manzagol, ``Extracting and
  composing robust features with denoising autoencoders,'' in \emph{ICML}, vol.
  307, 2008, pp. 1096--1103.

\bibitem{VAE_ICLR14}
D.~P. Kingma and M.~Welling, ``Auto-encoding variational bayes,'' in
  \emph{ICLR}, 2014.

\bibitem{VGAE_nips16}
T.~N. Kipf and M.~Welling, ``Variational graph auto-encoders,'' in
  \emph{NeurIPS Workshop on Bayesian Deep Learning}, 2016.

\bibitem{VQVAE_nips19}
A.~Razavi, A.~van~den Oord, and O.~Vinyals, ``Generating diverse high-fidelity
  images with {VQ-VAE-2},'' in \emph{NIPS}, 2019, pp. 14\,837--14\,847.

\bibitem{PIRL_cvpr20}
I.~Misra and L.~van~der Maaten, ``Self-supervised learning of pretext-invariant
  representations,'' in \emph{CVPR}, 2020, pp. 6706--6716.

\bibitem{GANs_iclr16}
A.~Radford, L.~Metz, and S.~Chintala, ``Unsupervised representation learning
  with deep convolutional generative adversarial networks,'' in \emph{ICLR},
  2016.

\bibitem{SGL_sigir21}
J.~Wu, X.~Wang, F.~Feng, X.~He, L.~Chen, J.~Lian, and X.~Xie, ``Self-supervised
  graph learning for recommendation,'' \emph{CoRR}, vol. abs/2010.10783, 2020.

\bibitem{MHCN_www21}
J.~Yu, H.~Yin, J.~Li, Q.~Wang, N.~Q.~V. Hung, and X.~Zhang, ``Self-supervised
  multi-channel hypergraph convolutional network for social recommendation,''
  in \emph{WWW}, 2021, pp. 413--424.

\bibitem{S3Rec_cikm20}
K.~Zhou, H.~Wang, W.~X. Zhao, Y.~Zhu, S.~Wang, F.~Zhang, Z.~Wang, and J.~Wen,
  ``S3-rec: Self-supervised learning for sequential recommendation with mutual
  information maximization,'' in \emph{CIKM}.\hskip 1em plus 0.5em minus
  0.4em\relax {ACM}, 2020, pp. 1893--1902.

\bibitem{BiGI_wsdm21}
J.~Cao, X.~Lin, S.~Guo, L.~Liu, T.~Liu, and B.~Wang, ``Bipartite graph
  embedding via mutual information maximization,'' in \emph{WSDM}.\hskip 1em
  plus 0.5em minus 0.4em\relax {ACM}, 2021, pp. 635--643.

\bibitem{DHCN_aaai21}
X.~Xia, H.~Yin, J.~Yu, Q.~Wang, L.~Cui, and X.~Zhang, ``Self-supervised
  hypergraph convolutional networks for session-based recommendation,'' in
  \emph{AAAI}, 2021, pp. 4503--4511.

\bibitem{HuKV08_icdm08}
Y.~Hu, Y.~Koren, and C.~Volinsky, ``Collaborative filtering for implicit
  feedback datasets,'' in \emph{ICDM}, 2008, pp. 263--272.

\bibitem{cdae_wsdm16}
Y.~Wu, C.~DuBois, A.~X. Zheng, and M.~Ester, ``Collaborative denoising
  auto-encoders for top-n recommender systems,'' in \emph{WSDM}.\hskip 1em plus
  0.5em minus 0.4em\relax {ACM}, 2016, pp. 153--162.

\bibitem{LFM_09}
Y.~Koren, R.~Bell, and C.~Volinsky, ``Matrix factorization techniques for
  recommender systems,'' \emph{Computer}, vol.~42, no.~8, pp. 30--37, 2009.

\bibitem{NCF_www17}
X.~He, L.~Liao, H.~Zhang, L.~Nie, X.~Hu, and T.~Chua, ``Neural collaborative
  filtering,'' in \emph{WWW}, 2017, pp. 173--182.

\bibitem{CML_www17}
C.~Hsieh, L.~Yang, Y.~Cui, T.~Lin, S.~J. Belongie, and D.~Estrin,
  ``Collaborative metric learning,'' in \emph{WWW}, 2017, pp. 193--201.

\bibitem{A3NCF_ijcai18}
Z.~Cheng, Y.~Ding, X.~He, L.~Zhu, X.~Song, and M.~S. Kankanhalli,
  ``A{\^{}}3ncf: An adaptive aspect attention model for rating prediction,'' in
  \emph{IJCAI}, 2018, pp. 3748--3754.

\bibitem{Chen_NARRE_www2018}
C.~Chen, M.~Zhang, Y.~Liu, and S.~Ma, ``Neural attentional rating regression
  with review-level explanations,'' in \emph{WWW}, 2018, pp. 1583--1592.

\bibitem{enmf_tois20}
C.~Chen, M.~Zhang, Y.~Zhang, Y.~Liu, and S.~Ma, ``Efficient neural matrix
  factorization without sampling for recommendation,'' \emph{{ACM} Trans. Inf.
  Syst.}, vol.~38, no.~2, pp. 14:1--14:28, 2020.

\bibitem{Pareto_86}
G.~Box and D.~Meyer, ``An analysis for unreplicated fractional factorials,''
  \emph{Technometrics}, vol.~28, pp. 11--18, 02 1986.

\bibitem{EHCF_aaai20}
C.~Chen, M.~Zhang, Y.~Zhang, W.~Ma, Y.~Liu, and S.~Ma, ``Efficient
  heterogeneous collaborative filtering without negative sampling for
  recommendation,'' in \emph{AAAI}, 2020, pp. 19--26.

\bibitem{KingmaB14}
D.~P. Kingma and J.~Ba, ``Adam: {A} method for stochastic optimization,'' in
  \emph{ICLR}, 2015.

\end{thebibliography}
	\bibliographystyle{IEEEtran}
	
	\vspace{-6mm}

\end{document}